\documentclass[12pt]{emulateapj}

\shorttitle{GRB\,060927 at $z = 5.47$}
\shortauthors{Ruiz-Velasco et al.}

\begin{document}

\title{Detection of GRB~060927 at \symbol{122} = 5.47: Implications for the Use of
Gamma-Ray Bursts as Probes of the End of the Dark Ages%
\altaffilmark{1}}
\author{
A. E. Ruiz-Velasco\altaffilmark{2,3}, 
H. Swan\altaffilmark{4},
E. Troja\altaffilmark{5,6,7},
D. Malesani\altaffilmark{2},
J. P. U. Fynbo\altaffilmark{2},
R. L. C. Starling\altaffilmark{5},
D. Xu\altaffilmark{2},
F. Aharonian\altaffilmark{8},
C. Akerlof\altaffilmark{4},
M. I. Andersen\altaffilmark{9},
M. C. B. Ashley\altaffilmark{10},
S. D. Barthelmy\altaffilmark{11},
D. Bersier\altaffilmark{12},
J. M. Castro Cer\'on\altaffilmark{2},
A. J. Castro-Tirado\altaffilmark{13},
N. Gehrels\altaffilmark{11},
E. G\"o\u{g}\"u\c{s}\altaffilmark{14},
J. Gorosabel\altaffilmark{13},
C. Guidorzi\altaffilmark{12,15},
T. G\"{u}ver\altaffilmark{16},
J. Hjorth\altaffilmark{2},
D. Horns\altaffilmark{8},
K. Y. Huang\altaffilmark{17},
P. Jakobsson\altaffilmark{18},
B. L. Jensen\altaffilmark{2},
\"{U}. K{\i}z{\i}lo\v{g}lu\altaffilmark{19},
C. Kouveliotou\altaffilmark{20},
H. A. Krimm\altaffilmark{11,21},
C. Ledoux\altaffilmark{22},
A. J. Levan\altaffilmark{23},
T. Marsh\altaffilmark{23},
T. McKay\altaffilmark{4},
A. Melandri\altaffilmark{12},
B. Milvang-Jensen\altaffilmark{2},
C. G. Mundell\altaffilmark{12},
P. T. O'Brien\altaffilmark{5},
M. \"{O}zel\altaffilmark{24},
A. Phillips\altaffilmark{10},
R. Quimby\altaffilmark{25},
G. Rowell\altaffilmark{8},
W. Rujopakarn\altaffilmark{26},
E. S. Rykoff\altaffilmark{4},
B. E. Schaefer\altaffilmark{27},
J. Sollerman\altaffilmark{2},
N. R. Tanvir\altaffilmark{5},
C. C. Th\"one\altaffilmark{2},
Y. Urata\altaffilmark{28},
W. T. Vestrand\altaffilmark{29},
P. M. Vreeswijk\altaffilmark{22},
D. Watson\altaffilmark{2},
J. C. Wheeler\altaffilmark{25},
R. A. M. J. Wijers\altaffilmark{30},
J. Wren\altaffilmark{29},
S. A. Yost\altaffilmark{4},
F. Yuan\altaffilmark{4},
M. Zhai\altaffilmark{31} and 
W. K. Zheng\altaffilmark{31}.
}

\altaffiltext{1}{Partly based on observations carried out with the ESO telescopes under
programmes 077.D-0661, 077.A-0667, 078.D-0416, and the large programme 177.A-f0591.}

\altaffiltext{2}{Dark Cosmology Centre, Niels Bohr Institute, University of Copenhagen, Juliane Maries Vej 30, 2100 Copenhagen, Denmark}

\altaffiltext{3}{Departmento de Astronom\'ia, Universidad de Guanajuato, Apartado Postal 144, 36000 Guanajuato, Mexico}

\altaffiltext{4}{University of Michigan, 2477 Randall Laboratory, 450 Church St., Ann Arbor, MI, 48109, USA}

\altaffiltext{5}{Department of Physics and Astronomy, University of Leicester, University Road, Leicester LE1 7RH, UK}

\altaffiltext{6}{INAF-Istituto di Astrofisica Spaziale e Fisica Cosmica Sezione di Palermo, Via Ugo La Malfa 153, 90146 Palermo, Italy}

\altaffiltext{7}{Dipartimento di Scienze Fisiche ed Astronomiche, Sezione di Astronomia, Universit\`a di Palermo, Piazza del Parlamento 1, 90134 Palermo, Italy}

\altaffiltext{8}{Max-Planck-Institut f\"{u}r Kernphysik, Saupfercheckweg 1, 69117 Heidelberg, Germany}

\altaffiltext{9}{Astrophysikalisches Institut, 14482 Potsdam, Germany}

\altaffiltext{10}{School of Physics, Department of Astrophysics and Optics, University of New South Wales, Sydney, NSW 2052, Australia}

\altaffiltext{11}{NASA/Goddard Space Flight Center, Greenbelt, MD 20771, USA}

\altaffiltext{12}{Astrophysics Research Institute, Liverpool John Moores University, Twelve Quays House, Egerton Wharf, Birkenhead CH41 1LD, UK}

\altaffiltext{13}{Instituto de Astrof\'isica de Andaluc\'ia (IAA-CSIC), Apartado de Correos, 3.004, E-18.080 Granada, Spain}

\altaffiltext{14}{Sabanc{\i} University, Orhanl{\i}-Tuzla 34956 Istanbul, Turkey}

\altaffiltext{15}{INAF-Osservatorio Astronomico di Brera, via Bianchi 46, 23807, Merate, Italy}

\altaffiltext{16}{Istanbul University, Science Faculty Department of Astronomy \& Space Sciences, Istanbul 34119, Turkey}

\altaffiltext{17}{Institute of Astronomy, National Central University, Ching-Li 32054, Taiwan}

\altaffiltext{18}{Centre for Astrophysics Research, University of Hertfordshire, College Lane, Hatfield, Herts AL10 9AB, UK}

\altaffiltext{19}{Middle East Technical University, 06531 Ankara, Turkey}

\altaffiltext{20}{NASA Marshall Space Flight Center, NSSTC, VP-62, 320 Sparkman Drive, Huntsville, AL 35805, USA}

\altaffiltext{21}{Universities Space Research Association, 10211 Wincopan Circle, Suite 500, Columbia, MD 21044, USA}

\altaffiltext{22}{European Southern Observatory, Alonso de C\'ordova 3107, Casilla 19001, Vitacura, Santiago 19, Chile}

\altaffiltext{23}{Department of Physics, University of Warwick, Coventry, CV4 7AL, UK}

\altaffiltext{24}{\c{C}anakkale Onsekiz Mart \"{U}niversitesi, Terzio\v{g}lu 17020, \c{C}anakkale, Turkey}

\altaffiltext{25}{Department of Astronomy, University of Texas, Austin, TX 78712, USA}

\altaffiltext{26}{Steward Observatory, University of Arizona, Tucson, AZ 85721, USA}

\altaffiltext{27}{Department of Physics and Astronomy, Louisiana State University, Baton Rouge Louisiana 70803, USA}

\altaffiltext{28}{Department of Physics, Saitama University, Shimo-Okubo, Sakura, Saitama 338-8570, Japan}

\altaffiltext{29}{Los Alamos National Laboratory, NIS-2 MS D436, Los Alamos, NM 87545, USA}

\altaffiltext{30}{Institute of Astronomy ``Anton Pannekoek'', University of Amsterdam, Kruislaan 403, 1098 SJ Amsterdam, The Netherlands}

\altaffiltext{31}{National Astronomical Observatories, Chinese Academy of Sciences, Beijing 100012, China}

\begin{abstract}

We report on follow-up observations of the gamma-ray burst GRB\,060927
using the robotic ROTSE-IIIa telescope and a suite of larger aperture 
ground-based telescopes. An optical afterglow was detected 20~s after the
burst, the earliest rest-frame detection of optical emission from any GRB.
Spectroscopy performed with the VLT about 13 hours after the trigger shows a
continuum break at $\lambda \approx 8070$~\AA, produced by neutral hydrogen
absorption at $z \approx 5.6$. We also detect an absorption line at 8158 \AA{}
which we interpret as \ion{Si}{2} $\lambda$ 1260 at $z=5.467$. Hence,
GRB\,060927 is the second most distant GRB with a spectroscopically measured
redshift. The shape of the red wing of the spectral break can be fitted by a
damped Ly$\alpha$ profile with a column density with $\log(N_{\rm
HI}$/cm$^{-2}) = 22.50 \pm 0.15$. We discuss the implications of this work for the
use of GRBs as probes of the end of the dark ages and draw three main
conclusions: {\it i)} GRB afterglows originating from $z \gtrsim 6$ should be
relatively easy to detect from the ground, but rapid near-infrared monitoring is
necessary to ensure that they are found; 
{\it ii)} The presence of large \ion{H}{1} column
densities in some GRBs host galaxies at $z>5$ makes
the use of GRBs to probe the reionization epoch via spectroscopy of the red
damping wing challenging; {\it iii)} GRBs  appear crucial to locate typical
star-forming galaxies at $z > 5$ and therefore the type of galaxies responsible
for the reionization of the universe.

\end{abstract}

\keywords{gamma rays: bursts (\objectname{GRB\,060927}) --- cosmology}

\section{INTRODUCTION}

It is well established that most long-duration gamma-ray bursts (GRBs) are
caused by the death of massive stars \citep[e.g.][]{galama,jensNature,stanek03} and,
due to their brightness, they can be seen throughout the observable universe
\citep{Kawai}. Given these facts, it has long been realized that GRBs could be
powerful probes of star-formation activity throughout the history of the
universe \citep[e.g.][]{wijers98}. The currently operating \textit{Swift}
satellite \citep{gehrels} has revolutionized the GRB field: it has not only
increased the detection rate of rapidly well-localized GRBs by roughly an order
of magnitude compared to previous missions, but it also detects much fainter
and more distant bursts \citep{palli06}.
In the \textit{Swift} era, GRBs have indeed become cosmological probes of the
early Universe, as already extensively predicted in the literature
\citep[e.g.][]{wijers98,lamb2000}. In particular, \citet{barkana} argue that
GRBs may be optimal probes of the epoch of reionization.

Here we present observations of GRB\,060927 for which we determine a very high
redshift ($z=5.467$, \S~\ref{sec:redshift}). In \S~\ref{sec:afterglow} we
discuss the light curve and the properties of the afterglow in the context of
the fireball model. In \S~\ref{sec:darkages} we end by revisiting the question
of how GRBs may be used to probe the end of the dark ages.

Throughout this paper we assume a cosmology with $H_0 = 70$
km~s$^{-1}$~Mpc$^{-1}$, $\Omega_{\rm m} = 0.3$ and $\Omega_\Lambda = 0.7$. In
this model, a redshift of 5.47 corresponds to a luminosity distance
$D_{\rm{lum}} = 51.9$ Gpc and a distance modulus $\mu = 48.6$~mag. At that distance,
$1\arcsec$ on the sky corresponds to 6.01 proper kpc and the look-back time is
12.4 Gyr (roughly 92\% of the time since the Big Bang).

\section{OBSERVATIONS AND DATA ANALYSIS}

\subsection{High-Energy Properties}
\label{sec:highenergy}

GRB\,060927 and its X-ray afterglow were detected by the Burst Alert Telescope
(BAT) and the X-ray Telescope (XRT) on-board the \textit{Swift} spacecraft on
2006 September 27.58860 UT. At the measured redshift of $z=5.47$
(\S~\ref{sec:redshift}) the BAT 15--150~keV energy band and the XRT 0.3--10~keV
energy band correspond to 97--970~keV and 1.9--64.7~keV in the source rest
frame, respectively. Here we describe the burst phenomenology as observed by
the \textit{Swift} instruments. Data have been analyzed using the standard
analysis software distributed within FTOOLS%
\footnote{\texttt{http://heasarc.gsfc.nasa.gov}\,.} v.~6.1.1.

The gamma-ray prompt emission shows a complex light curve with an initial
bright episode, split into two peaks. The peak flux is reached during the first
peak, $\sim 0.7$~s after the trigger ($T_0$), and the emission subsequently
decays to the background level at $\sim T_{0}+9$~s. A fainter bump is visible
at $\sim T_{0}+20$~s. The observed duration is $T_{90} = 22.6\pm0.3$~s, which,
accounting for the $(1+z)$ time stretching factor, corresponds to
$3.54\pm0.05$~s in the source rest frame. GRB\,060927 is thus a long-duration
GRB \citep{Kouv93}, both in the observer's frame and in the rest frame.

A simple power-law model is not a good fit to the BAT time-average spectrum of
the burst ($\chi^2 = 72$ for 57 d.o.f.). A significant improvement is achieved
by adopting a cutoff power law ($\chi^2 = 58$ for 56 d.o.f.), which provides a
photon index $\Gamma = 0.9 \pm 0.4$ and a peak energy $E_{\rm p} =
72_{-6}^{+39}$~keV. The observed burst fluence is $1.1^{+0.2}_{-0.7} \times
10^{-6}$ erg~cm$^{-2}$ in the 15--150~keV band. A fit with a Band model
\citep{band93} does not significantly improve the $\chi^2$ value, albeit this
can be due to the low sensitivity of BAT at high energies. Using the best-fit model,
we can classify GRB\,060927 as an X-ray rich GRB using the definition proposed
by \citet{Lamb}. During the initial prompt emission, hard-to-soft spectral
evolution is present, the first spike being significantly harder than the
second. The spectrum of the late bump is consistent with that of the second
spike.

The X-ray afterglow was detected by XRT about 70~s after the trigger and it was
monitored for the following 3 days. The X-ray afterglow position, calculated
using the updated boresight \citep{gcn5750}, is $\mbox{R.A. (J2000)} = 21^{\rm
h} 58^{\rm m} 12\fs03$, $\mbox{decl. (J2000)} = 05\arcdeg 21\arcmin 49\farcs4$,
with a 90\% error radius of 3\farcs7. This position is within 0\farcs7 of the
optical counterpart (\S~\ref{sec:imaging}). The X-ray light curve displays an
initial shallow decay with a slope $\alpha_1 = 0.71 \pm 0.06$, steepening to
$\alpha_2 = 1.35\pm0.11$ at $\sim4$~ks after the trigger. The spectrum is
modeled with an absorbed power law with photon index $\Gamma = 1.87\pm0.17$.
After correcting for the Galactic absorption $N_{\rm H} = 5.2 \times
10^{20}$~cm$^{-2}$, the 90\% confidence limit for the intrinsic absorption is
$N_{\rm H} < 3.4 \times 10^{20}$~cm$^{-2}$ (observer frame). Indeed, little
observed X-ray absorption is expected at this large redshift (\citealt{grupe};
although see GRB\,050904: \citealt{Watson06,Cusumano06,Campana07}), since most
of the important absorption edges have been redshifted out of the observed
X-ray region.
The unabsorbed flux in the 0.3--10~keV energy band from $T_0+75$~s to
$T_0+11.3$~ks is $(5.7\pm0.8) \times 10^{-12}$~erg~cm$^{-2}$~s$^{-1}$ from
which we derive a lower limit of $10^{-6}$~erg~cm$^{-2}$ to the fluence emitted
by the X-ray afterglow.

\subsection{Ground-Based Imaging Observations}
\label{sec:imaging}

The optical afterglow was first detected by the 45~cm telescope of the Robotic
Optical Transient Search Experiment in Australia (ROTSE-IIIa). The first
unfiltered image started 16.5~s after the burst trigger \citep{schaefer},
which corresponds to 3~s in the GRB rest frame. Therefore, this is the
earliest rest-frame detection of optical emission from any GRB%
\footnote{See \texttt{http://grad40.as.utexas.edu/grblog.php}\,.}.

Subsequent follow-up observations were obtained by several ground-based
facilities. In particular, we used the data from the following instruments: 
the robotic 2.0~m Faulkes Telescope South (FTS) located at the Siding Spring
Observatory, Australia, equipped with RatCam;
the 105~cm Schmidt Telescope at the Kiso Observatory in Japan;
the 2.56~m Nordic Optical Telescope (NOT), equipped with ALFOSC at La Palma
(Canary Islands); the 4.2~m William Herschel Telescope (WHT), using the Aux
port imager, also at La Palma; the 1.54~m Danish Telescope situated at La
Silla Observatory in Chile; and the 8.2~m Antu and Kueyen Unit Telescopes of
the ESO Very Large Telescope (VLT) on Cerro Paranal in Chile, equipped with
FORS2 and FORS1, respectively. Near-infrared (NIR) observations were secured
using the ISAAC camera on the ESO VLT Antu. In Table~\ref{tb:log}, we provide
the complete observing log.
The reduction of the optical and NIR data was performed using standard IRAF%
\footnote{IRAF is distributed by the National Optical Astronomy Observatories,
which are operated by the Association of Universities for Research in
Astronomy, Inc., under cooperative agreement with the National Science
Foundation.}
tools. The photometry of the afterglow was performed using both PSF-matched and
aperture photometry.

Using the VLT image from 2006 September 30, we determined the position of the 
afterglow to be $\mbox{R.A. (J2000)} = 21^{\rm h} 58^{\rm m} 12\fs02$, 
$\mbox{decl. (J2000)} = +05\arcdeg 21\arcmin 48\farcs9$, relative
to $\approx 30$ isolated, non saturated USNO-B1 stars. The estimated uncertainty
is about 0\farcs2 in each coordinate.

\begin{figure}
\centering
\plotone{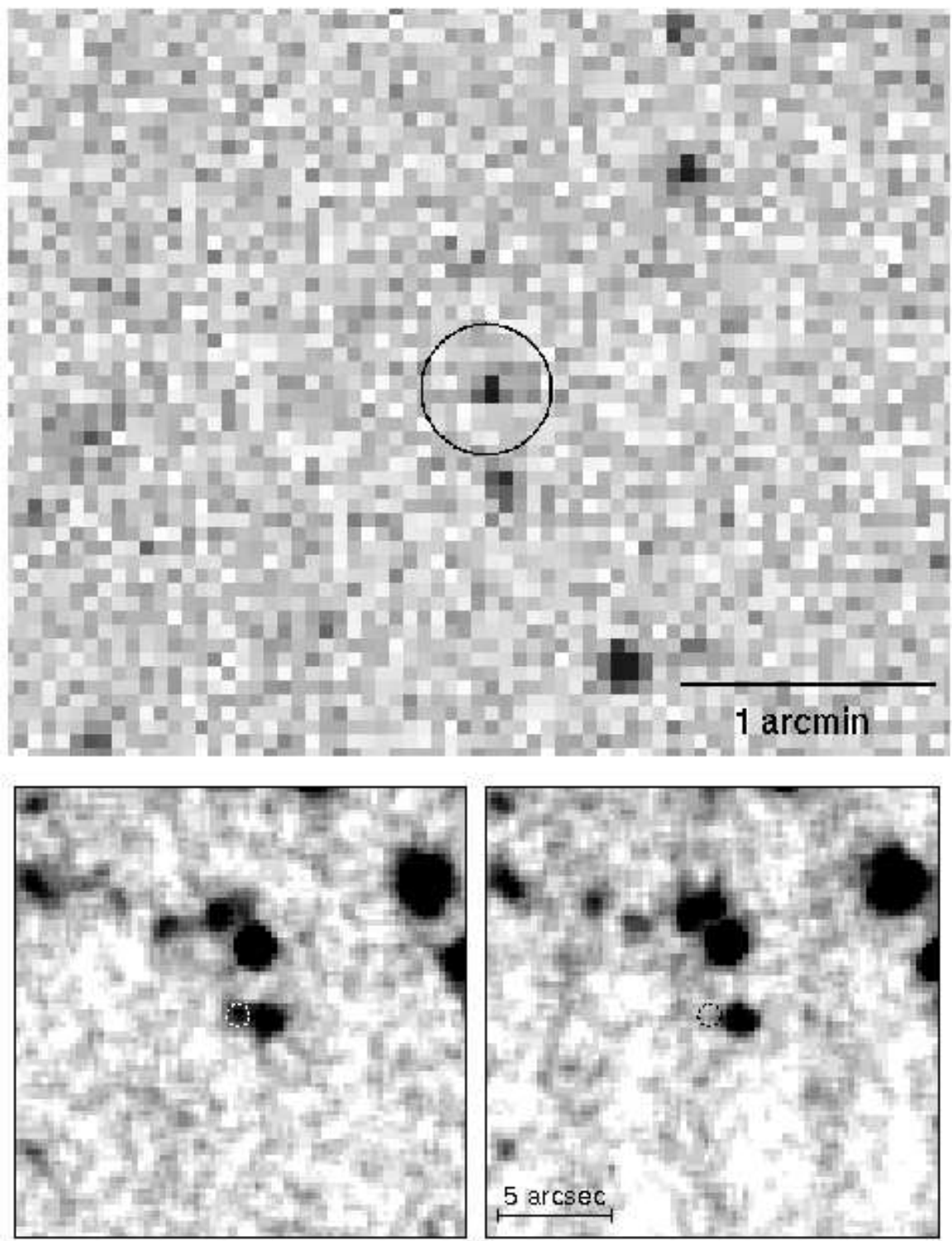}
\caption{Top: unfiltered image from ROTSE-IIIa obtained 19~s after the
trigger. Bottom: VLT $I$-band images 2.6 days (left) and 15.5 days (right)
after the trigger. The circles indicate the position of the optical
afterglow. North is up, East is left.\label{fg:image}}
\end{figure}

The detection of the afterglow by ROTSE and the lack of an optical counterpart
observation by UVOT on board \textit{Swift} \citep{Oates06a} revealed within a
few minutes that the afterglow of GRB\,060927 was very red. This was soon
confirmed by the FTS observations, which showed an $I$-band afterglow with no
corresponding $R$-band emission within the first 30 minutes after the GRB
\citep{guidorzia}. Later, deep WHT observations taken 8.9~hr after
the burst indicated $R-I > 2.0$. The $R$-band image from the NOT taken 6.2~hr
after the burst reveals a low signal-to-noise detection of the afterglow with $R \sim 24.4$.
After realizing the high redshift nature of the GRB (see below) we obtained
deep $I$-band imaging with the VLT leading to a detection of the afterglow
$\sim 2.6$ days after the burst, and an upper limit at 15 days. We also
obtained VLT/ISAAC NIR imaging leading to tentative detections in the $J$ and
$K$ bands. No host galaxy was detected down to $I > 25.8$ ($3\sigma$ limit). In
Fig.~\ref{fg:image} we show the afterglow of GRB\,060927 as detected in the
ROTSE and $I$-band VLT exposures.

\subsection{Spectroscopic Observations and Redshift Determination}
\label{sec:redshift}

Spectroscopic observations were performed with the FORS1 instrument at the VLT
12.5~hr after the trigger. We used the grism 300V covering the spectral range
from 3600 to 8900~\AA{} at a resolution of 12~\AA{} (using a 1\farcs0 wide
slit). The acquisition was performed by aligning the slit with a galaxy
$3\farcs1$  North of the afterglow and using a position angle for which the
slit covered the afterglow position ($\mathrm{PA} = 20\arcdeg$ West of North).
The afterglow was not detected in the $R$-band acquisition image down to a
limiting magnitude $R \sim 24.0$.

\begin{figure}
\centering
\plotone{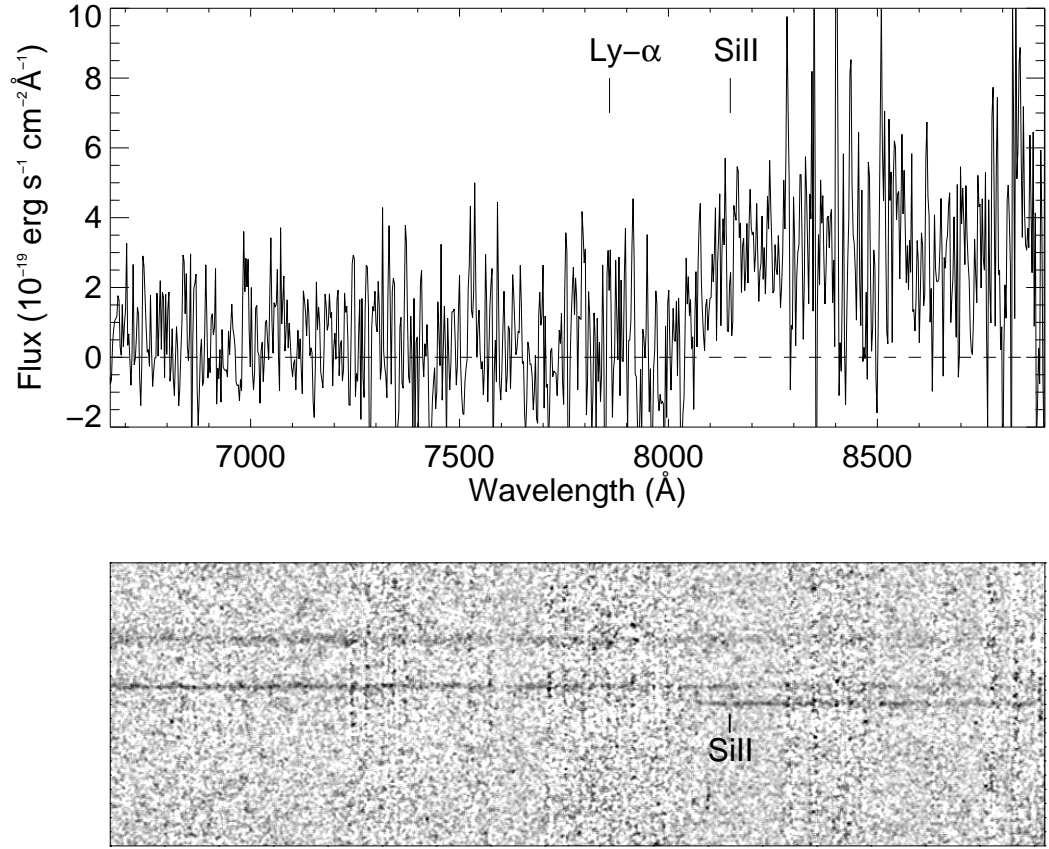}
\caption{Afterglow spectrum of GRB\,060927 taken 12.5~hr after the trigger.
The one-dimensional spectrum is shown in the upper panel with the identified
absorption Ly$\alpha$ and \ion{Si}{2} lines indicated. The CCD image of the spectrum in
the corresponding wavelength range is shown in the lower panel (bottom trace).\label{fg:spec}}
\end{figure}

In the spectrum, a faint continuum is detected at the expected position of the
afterglow, but only redward of $\lambda = 8070$~\AA. This feature is similar
to what is seen in the spectra of quasars at redshift $z > 5$ \citep{cool}.
From this clear and sharp discontinuity in the spectrum we conclude that
GRB\,060927 is a high-redshift event, and is not red because of dust extinction. We
interpret the absorption blueward of the break, where the flux level is
consistent with zero, as a Ly$\alpha$ forest trough at redshift $z \approx
5.6$. In the sky-subtracted two-dimensional spectrum we also clearly detect a
single absorption line at $\lambda = 8148$~\AA{} in a wavelength region clear of bright sky
lines (see Fig.~\ref{fg:spec}). This is most likely \ion{Si}{2} $\lambda$ 1260
at a redshift of $z = 5.467$. At this redshift we would also expect
\ion{O}{1}/\ion{Si}{2} at 1303~\AA{} and \ion{C}{2} at 1335~\AA{}, but these
fall on bright sky lines and hence would not be detectable. This value for the
redshift supersedes our preliminary determination \citep{Fynbo06}. The difference in
the redshift derived using the hydrogen and silicon lines is due to the
presence of strong neutral hydrogen absorption (see \S~\ref{sec:Lyalpha}): the
damping wing of the Ly$\alpha$ trough effectively shifts the cutoff at redder
wavelengths, mimicking the effect of a larger redshift. In Fig.~\ref{fg:spec}
we show the one-dimensional flux- and wavelength-calibrated spectrum (upper
panel) and the processed CCD image (lower panel). In the latter, we can see the
sharp continuum break in the afterglow spectrum. The two traces above
correspond to nearby objects falling on the slit. We last note that the
pseudo-redshift for this burst \citep{pseudoz_GCN,pseudoz_paper} is $\hat{z} =
2.37 \pm 0.75$, significantly different (4$\sigma$) from the spectroscopic
value.

\subsection{Fitting the Red Damping Wing}
\label{sec:Lyalpha}

\begin{figure}
\centering
\plotone{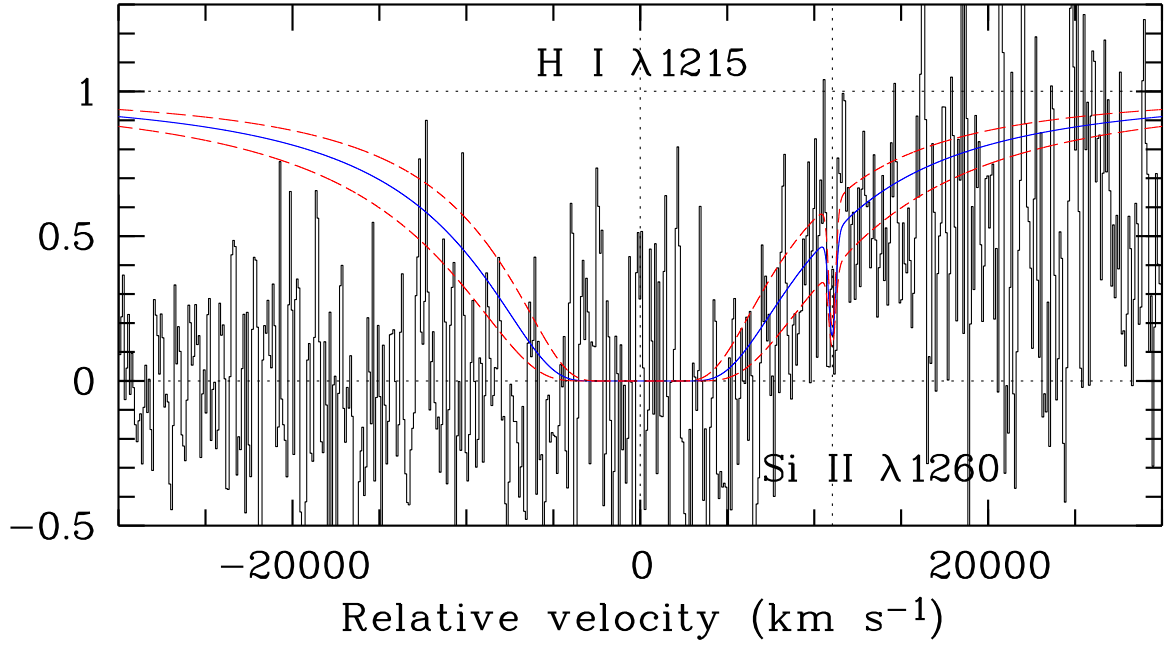}
\caption{Afterglow spectrum with the model fit to the Ly$\alpha$ and
\ion{Si}{2} $\lambda$ 1260 absorption line profiles (solid curve). The
corresponding column density has $\log(N_{\rm HI}/{\rm cm}^{-2}) =
22.50 \pm 0.15$.\label{fg:Lyalpha}}
\end{figure}

Measuring the shape of the absorption profile of the damping wing provides a
determination of the column density of the neutral hydrogen along the line of
sight to the progenitor in the host \citep{madau}. Using the Fit-Lyman program
within MIDAS, we fitted the wing shape with a Damped Ly$\alpha$ absorber system
centered at 7861~\AA{}, assuming $z_{\rm DLA} = z_{\rm Si\,II} = 5.467$. In
Fig.~\ref{fg:Lyalpha} we show the fit to the Ly$\alpha$ and \ion{Si}{2}
$\lambda$ 1260 absorption line profiles in a $\pm 30,000$~km~s$^{-1}$ velocity
interval (solid curve). The best-fit column density is estimated to have
$\log(N_{\rm HI} / {\rm cm}^{-2}) = 22.50 \pm 0.15$.
The $N_{HI}$ value is firmly set by the core of the DLA line, i.e., the part of the 
spectrum which has zero residual flux, whatever the continuum normalisation is.
while the error is mostly due to
the continuum placement. The $N_{\rm HI}$ value is in the
range of those found for other \textit{Swift} GRBs at $z > 2$ \citep{palli06b},
although at the high end of the distribution.

\section{AFTERGLOW PROPERTIES}
\label{sec:afterglow}

\subsection{Light Curve}
\label{sec:lc}

A major complication with the analysis of the optical light curve is the
photometric calibration. The ROTSE data were taken with no filter. For the
other telescopes, the $I$-band data were obtained using three different filter
variants: an SDSS-like $i$ filter for the FTS, the cutoff Harris $I$ for the
WHT, and the ``standard'' Bessell $I$ for the VLT. For all of them, a
significant fraction of the light within the effective filter passband is
absorbed by the Ly$\alpha$ forest. This makes it difficult to obtain a
meaningful calibrated magnitude or flux from the measured counts.

The ROTSE telescope spectral response is
determined by the CCD characteristics, which results in an equivalent broadband
filter with a spectral window from 3000 to 10,000~\AA{} and a peak efficiency
around 6000~\AA{}. Due to absorption by the Ly$\alpha$ forest, we expect little
flux blueward of 8000~\AA{} and therefore our effective response for the
optical transient is closer to the $I$ band rather than to the $R$ band. To obtain a
consistently calibrated afterglow flux, we have used the SDSS photometry of
neighboring stars \citep{CoolGCN,Adelman} to establish the numerical
conversion factors from ROTSE magnitudes to photon count rates. Six bright,
unsaturated and isolated SDSS stars were selected near the GRB for this
calibration, spanning a broad range of stellar color. By using the four-band
$griz$ SDSS data for these stars, we estimated the total flux in the ROTSE
passband. After convolving these total fluxes with the ROTSE spectral response
function a model flux for each of the six comparison stars was produced. The RMS
variation of these computed values to the ROTSE measurements was 0.09~mag,
which gives a reasonable measure of the absolute accuracy of this procedure.
We note, however, that since the counterpart was close to detection threshold, 
statistical uncertainties were always dominating the total error budget.

The next step requires some assumptions about the spectral energy distribution
of the GRB afterglow. We adopted a ``typical'' power-law form, $F_\nu(\nu)
\propto \nu^{-\beta}$, with $\beta = 0.75$ (see also \S~\ref{sec:SED}). This
spectrum was folded first with the Ly$\alpha$ absorption in the intergalactic
medium (IGM) using the model described by \citet{Meiksin2005} and, second, with
the ROTSE spectral response. In the ROTSE passband, 83\% of the GRB flux is
absorbed by the IGM. Since optical observations must be compared over four
different optical passbands, it is desirable to pick a reference wavelength
$\lambda_0$ common to all of them in order to minimize the effects of the
uncertainty in $\beta$. A sensible choice for $\lambda_0$ requires it to be
higher than the Ly$\alpha$ cutoff frequency (corresponding to 8000~\AA) and
lower than the long-wavelength cutoff of any of the various $I$-band filters
(i.e., 8650~\AA{} for FTS $i$). From the above constraints, we chose
arbitrarily $\lambda_0 = 8190$~\AA. Rather than quoting magnitudes based on large
extrapolations to actual filter bandwidths, we computed the spectral flux
density at the value of $\lambda_0$ indicated above. We also checked the effect
of varying the afterglow spectral index $\beta$ by $\pm 0.25$. Since the
observed spectral region is narrowly confined by the Ly$\alpha$ cutoff and by
the edge of the filters, the corresponding fluxes are not much changed (by
$\sim 1\%$). The inferred flux numerical values are listed in
Table~\ref{tb:log}, and have been corrected for the Galactic extinction $E(B-V)
= 0.06$~mag \citep{Nh} as well.

A similar, but simpler procedure was carried out for the data obtained from the
other telescopes. These were in fact equipped with narrow-band filters, so 
the fraction of light inside the filter lost due to the
intergalactic medium absorption was computed. Taking into account the appropriate filter
sensitivity functions and the detector quantum efficiencies, this fraction is
$68\%$, $47\%$, and $56\%$ for the FTS, WHT, and VLT, respectively.

We also note that the afterglow was marginally detected in the $R$ band by the
NOT and by the Danish 1.54 m telescopes. This detection is very likely due to
the extension of the $R$-band filter response redward of 8000~\AA. Using the
tabulated filter efficiency, we estimate that $\approx 6\%$ of the light is
transmitted (3~mag suppression). Considering the observed color $R - I = 2.3
\pm 0.3$, and correcting for the Ly$\alpha$ suppression in both filters, we
infer an intrinsic $R - I = 0.2 \pm 0.3$, which is consistent with normal
afterglow colors given the large uncertainty. We thus confirm that very small
light is transmitted blueward of the Ly$\alpha$ cutoff.



\begin{figure}
\centering
\plotone{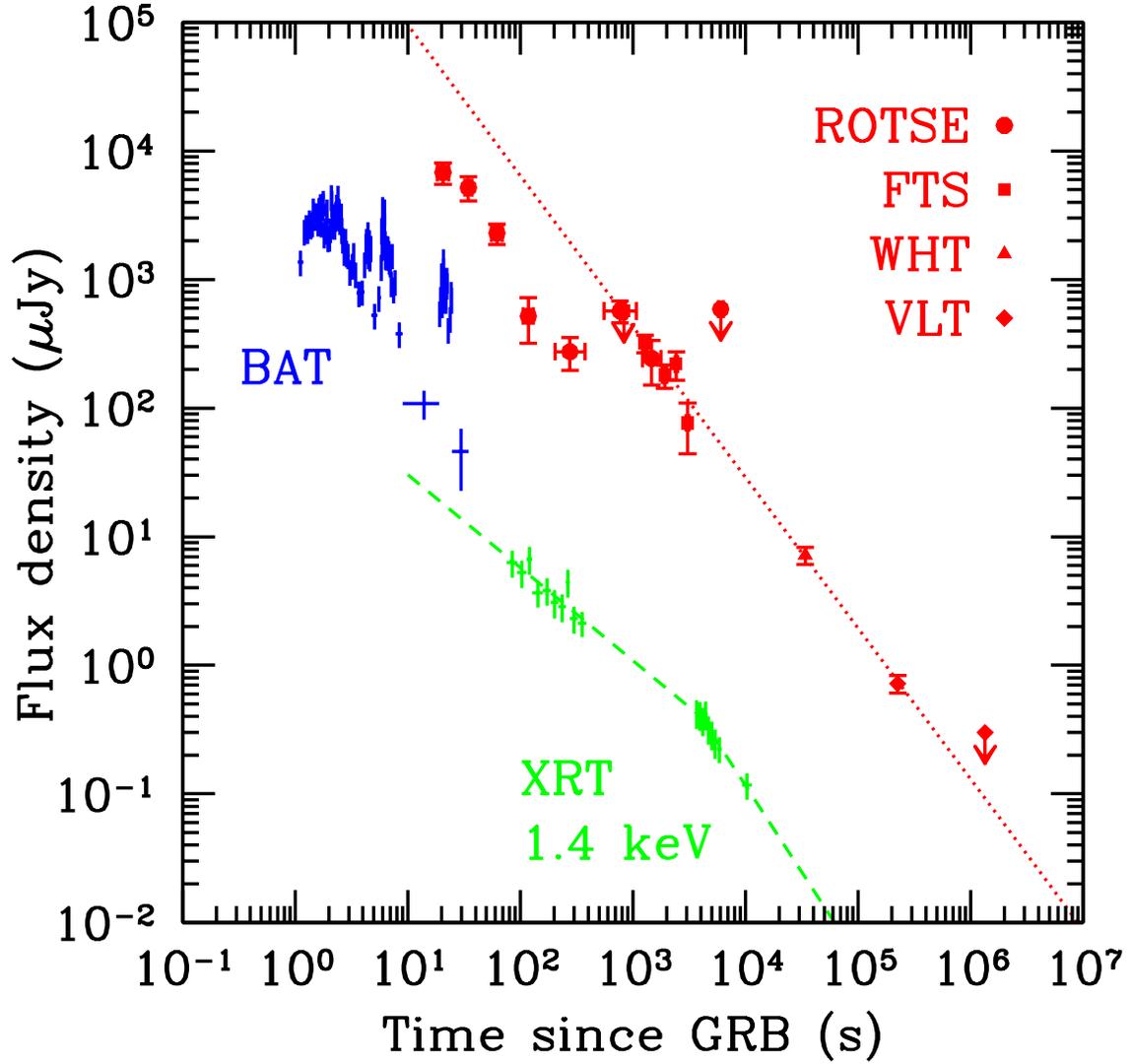}
\caption{Optical (8190~\AA, filled symbols) and X-ray (1.4 keV, crosses) light
curves of GRB\,060927. The early BAT measurements were extrapolated to the
XRT range using the best-fit spectral model. Optical data were corrected for
the Galactic extinction. The dotted line shows the best-fit power-law model
to the $I$-band light curve for $t > 500$~s, while the dashed line shows the
best-fit broken power-law model to the X-ray light curve.\label{fg:lc}}
\end{figure}

In Fig.~\ref{fg:lc} we present the light curve of the GRB\,060927 afterglow,
showing $I$-band corrected fluxes and X-ray data points. The X-ray afterglow
seems to be rather typical \citep{Zhang06,OB06,Nousek06}. At early times
($t \lesssim 3000 s$) it is characterized by a shallow decay, with temporal
index $\alpha_1 \approx 0.7$. Comparison with the extrapolation of the BAT flux also
implies a steep decline before the beginning of the observation, showing that
GRB\,060927 had all the three ``canonical'' phases. The shallow decline
steepened to $\alpha_2 \approx 1.3$ at $\sim 4$~ks after the trigger.

The situation is different in the optical. After $\sim 500$~s the data are
fairly well described by a single power law with decay index $\alpha = 1.17 \pm
0.03$ ($\chi^2 = 3.6$ with 6 d.o.f.). The early ROTSE data, however, lie
significantly below the extrapolation of the late light curve; a single
power-law fit to the full $I$-band data set provides an unacceptable $\chi^2 =
86$ for 11 d.o.f.. A flat light curve between 300 and 1000 s is also visible 
in the Kiso $R$-band data. The last ROTSE points agree well with the FTS data, so that
the difference cannot be due to an intercalibration problem. 

 One possibility is
that a break was present also in the optical, at $\sim 500$~s after the
trigger. Alternatively, the peak at $\sim 300$~s after the GRB may suggest a
rebrightening. The flux after the peak follows a single power law, indicating
that no flare was present. Rather, the flux was steadily larger than the
extrapolation of the behavior before the brightening. This step-like behavior
is consistent with being due to an episode of energy injection. Energy
injection is also often invoked to explain the shallow X-ray decay
\citep[e.g.][]{Zhang06}, albeit in GRB\,060927 the shallow phase ended significantly
later than the brightening episode. A different possibility is that the early
emission was due to another component, like a reverse shock or central engine
activity (note that the first ROTSE measurements are simultaneous to the last
peak of the prompt emission).


It is remarkable that the observed $I$-band flux is declining by the very early
observation (3 s in the GRB rest frame). This is difficult to explain in the
context of the standard afterglow model \citep[e.g.][]{Sari98}. The flux is
indeed expected to rise before the so-called deceleration time $t_{\rm dec}$.
If the early emission was due to the forward shock, our data constrain $t_{\rm
dec} \la 3$~s (GRB frame). This translates into a lower limit on the fireball
initial Lorentz factor $\Gamma_0 \ga 1000$. Furthermore, a break in the light
curve is expected when the injection frequency $\nu_{\rm inj}$ crosses the
observed band. The lack of such a break requires that $\nu_{\rm inj}$ was
already below the optical range at the beginning of our observations. Very small
equipartition parameters are required for this to happen. These difficulties
may also support the idea that the early emission was not due to the forward
shock.




The optical and X-ray decay slopes at late times are marginally consistent.
This is in agreement with the broad-band spectral analysis carried out at $t
\sim 5700$~s (\S~\ref{sec:SED}), which shows that the overall SED is well
described by a single power law. The hard spectral index $\beta \approx 0.7$
(Table~\ref{tb:SED}) suggests the cooling frequency was above the X-ray region.
However, as discussed before, at early times the X-ray emission was flatter,
and the optical light curve showed no steepening at the X-ray break time. This
behavior has already been noted in a number of \textit{Swift} bursts
\citep[e.g.][]{Panaitescu06}.



\subsection{Burst Energetics}
\label{sec:energetics}

The time-averaged spectral energy distribution of the prompt emission is well
fitted by a cutoff power law with a photon index $\Gamma = 0.9\pm0.4$ and a
peak energy $E_{\rm p} =72_{-6}^{+39}$~keV. In the burst rest frame, the
intrinsic peak energy was $E'_{\rm p} = 466_{-40}^{+250}$~keV. The burst
fluence over the $T_{90}$ interval in the 15--150 keV band was
$1.1_{-0.7}^{+0.2} \times 10^{-6}$~erg~cm$^{-2}$. This corresponds to an
isotropic-equivalent energy $E_{\rm iso} = 7.7^{+2.8}_{-5.0} \times
10^{52}$~erg in the rest-frame 1--10\,000~keV band. The values of $E'_{\rm p}$
and $E_{\rm iso}$ make this GRB consistent with the Amati relation
\citep{Ama02,Ama06}.


We can provide only a lower limit to any light curve break, which corresponds
to $t_{\rm b} \ga 2.6$~days (the epoch of our last detection). The lower limits
for the jet opening angle and for the beaming-corrected energy are $\vartheta >
5\arcdeg$ and $E_{\rm jet} > 2.6 \times 10^{50}$~erg, respectively. The
$E'_{\rm p}$ vs $E_{\rm jet}$ correlation \citep{Ghirla04} predicts for this
burst $t_{\rm b} = 10^{+8}_{-4}$~days \citep{Campana07}, thus GRB\,060927 is
consistent with this relation, although it does not constrain it (see also
\citealt{Ghirla07}).

\subsection{The Afterglow Spectral Energy Distribution}
\label{sec:SED}

\begin{figure}
\centering
\includegraphics[angle=-90,width=0.8\textwidth]{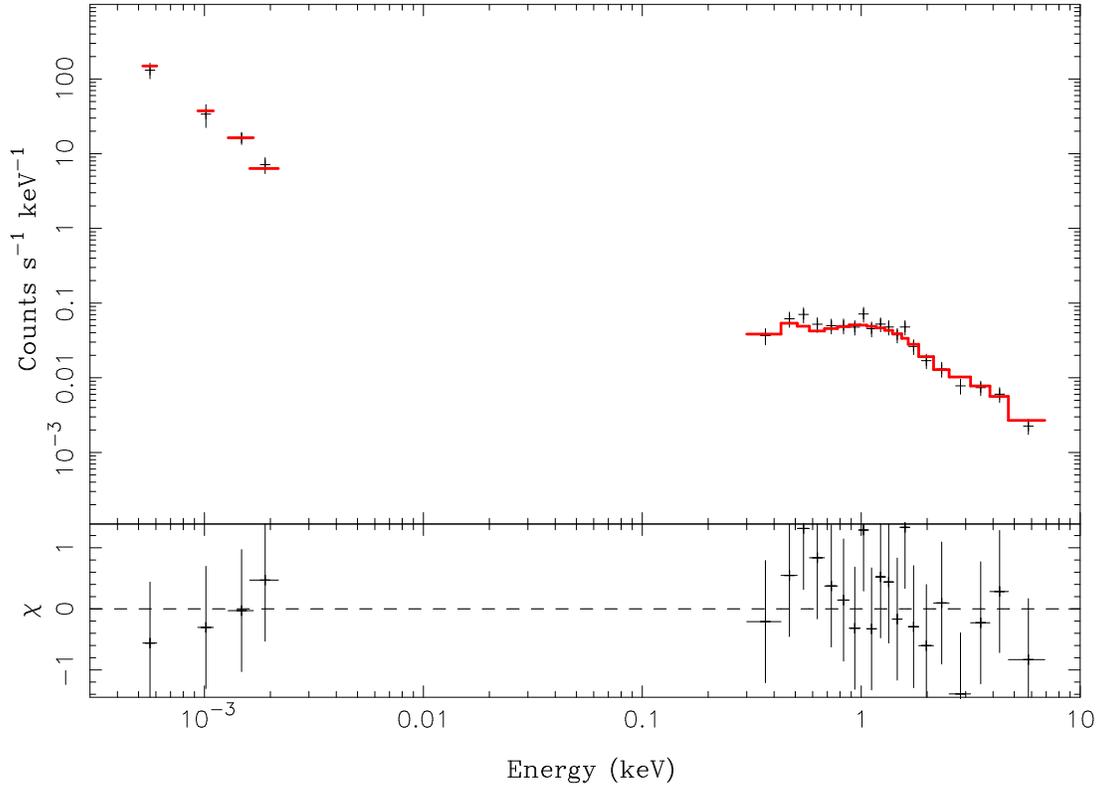}
\caption{Example of fit to the NIR/X-ray SED. The model (thick solid line) shows a single power law with Milky-Way extinction.
The lower panel shows the residuals of the fit.\label{fg:SED}}
\end{figure}

We have constructed the spectral energy distribution (SED) of the afterglow
covering from NIR to X-ray wavelengths at 5709~s after the \textit{Swift} BAT
trigger. Both a power law (PL) and a broken power law (BPL) have been fitted
including the effect of dust absorption, adopting the extinction curves of the
Milky Way (MW), the Large Magellanic Cloud (LMC) and the Small Magellanic Cloud
(SMC) as parametrized by \citet{pei}, and of photoelectric absorption in the
X-ray band (assuming Solar metallicity). All fits include a fixed Galactic
absorption component, with $N_{\rm H,Gal} = 5.2 \times 10^{20}$ cm$^{-2}$
\citep{dl90} and $E(B-V) = 0.062$~mag. Correction for Ly$\alpha$ blanketing
was applied to the $I$ and $R$ data. When excluding the $R$ band data, 
which are severely suppressed, the parameters did not change within the errors.
All the results are listed in 
Table~\ref{tb:SED}, and the best fit is shown in Fig.~\ref{fg:SED}.
Also considering the broad-band SED, we do not measure any
intrinsic X-ray absorption, consistent with the XRT analysis alone. A broken
power law with photon indices $\Gamma_1$ and $\Gamma_2$ provides no improvement
upon the single power law fits, and the break energies are difficult to
constrain. We also test for the presence of a cooling break in the observed
energy range by fixing $\Gamma_1 = \Gamma_2 - 0.5$, as expected from the
standard synchrotron theory, but found that no such break is required. The
optical absorption is found to be low, in agreement with many previous studies
of GRB host galaxy extinction \citep[e.g.,][]{Kann06,extincs}. However, it is
not possible in this case to differentiate between the three tested extinction
curves. The 2175~\AA{} bump characteristic of the MW extinction curve would
occur at $\sim 1.4$~$\mu$m when redshifted to $z = 5.47$, falling in between
the $J$ and $K$ bands where we have no observations. From our data we can,
however, obtain an accurate measurement of the power-law slope at this epoch,
which is consistent within the errors with the value derived from fitting the
X-ray data alone (\S~\ref{sec:highenergy}). The similar decay indices in the
optical and X-ray ranges also suggests that both components lie on the same
spectral segment.

\section{GAMMA-RAY BURSTS AS PROBES OF THE DARK AGES}
\label{sec:darkages}

Spectroscopy of GRB afterglows has been argued to be a powerful tool to probe
the red Ly$\alpha$ damping wing resulting from the neutral intergalactic medium
before reionization \citep{barkana}. In fact, they have simple power-law continuum
spectra and a relatively small ionizing effect on
their environments (compared to QSOs). This shape allows one to directly infer the neutral
hydrogen column and hence the state of reionization \citep{madau}. However, as
pointed out by \citet{totani06}, this method may be compromised by the strong
hydrogen absorption due to the dense environments that long GRBs usually seem
to be situated in \citep{palli06b}.
Our observation that such a high absorbing column is also present in a $z > 5$
GRB confirms that it is likely difficult to use afterglow spectra
as a diagnostic of reionization due to the Ly$\alpha$ damping wing. We also note,
however, that GRB-DLA column densities are usually lower than for GRB\,060927:
in the sample by \citet{palli06b} only two out of 18 events ($\approx 10\%$)
have $N_{\rm HI} > 10^{22.5}¯\mathrm{~cm}^{-2}$. Furthermore, as pointed
out by \citep{BrommLoeb07}, it is conceivable that higher-redshift bursts may
explode inside less massive halos, and thus with lower column densities.

The outlook for the use of GRBs as star-formation tracers is much better. For
example, it has also been suggested by \citet{wijers98} that GRBs could be
{\it ideal} tracers of the average star formation density due to their
brightness, independently of dust extinction and the fact that they originate
from a single (or double) stellar progenitor and so do not require a detectable
host galaxy \citep[see also][]{conselice05}. Evidence has been mounting that
GRBs may only trace star formation in relatively metal-poor environments
\citep[e.g.][]{lefloch,fynbo,Tanvir04,Soller05,fruchter,stanek06,modjaz07}. In
the  popular collapsar model, GRBs are expected to be formed only by stars with
metallicity below about 0.2--$0.3 Z_{\odot}$ \citep[][but see also
\citealt{fryer}]{heger,MacF,Hirschi,WH06,LangerNorman06}. However, before $z
\approx 5$ the majority of the star-formation should occur in environments with
$Z < 0.2 Z_{\odot}$ \citep{nuza}, and therefore the redshift distribution of
$z>5$ GRBs should directly map the star formation history of the universe back
to the first era of population-II objects, and possibly the first population-III
stars \citep[e.g.][]{bromm,belczinski}.

Furthermore, GRBs afterglows pinpoint their host galaxies, provide redshifts for them,
and in some cases measures of their metallicity, gas density and dynamical
state\footnote{Although the existence of the GRB might select enviroments with 
special properties, 
recent observations \citep{Chen06,Watson07,Vreeswijk07} have shown
that the material being probed by afterglow spectroscopy is not in the immediate surroundings
of the GRB site, implying that it may be representative of the host galaxy properties.}. 
This provides a route to identify and study the
types of galaxies that are responsible for the bulk of the star formation, and
hence for the bulk of the (re)-ionization of the universe. 
This is very important
since other methods to locate high redshift galaxies with current 
instrumentation only locate the brightest systems that are too rare to account
for the bulk of the ionization \citep{labbe07}, possibly with the exception of
gravitational lensing \citep{stark}.

\begin{figure}
\centering
\plotone{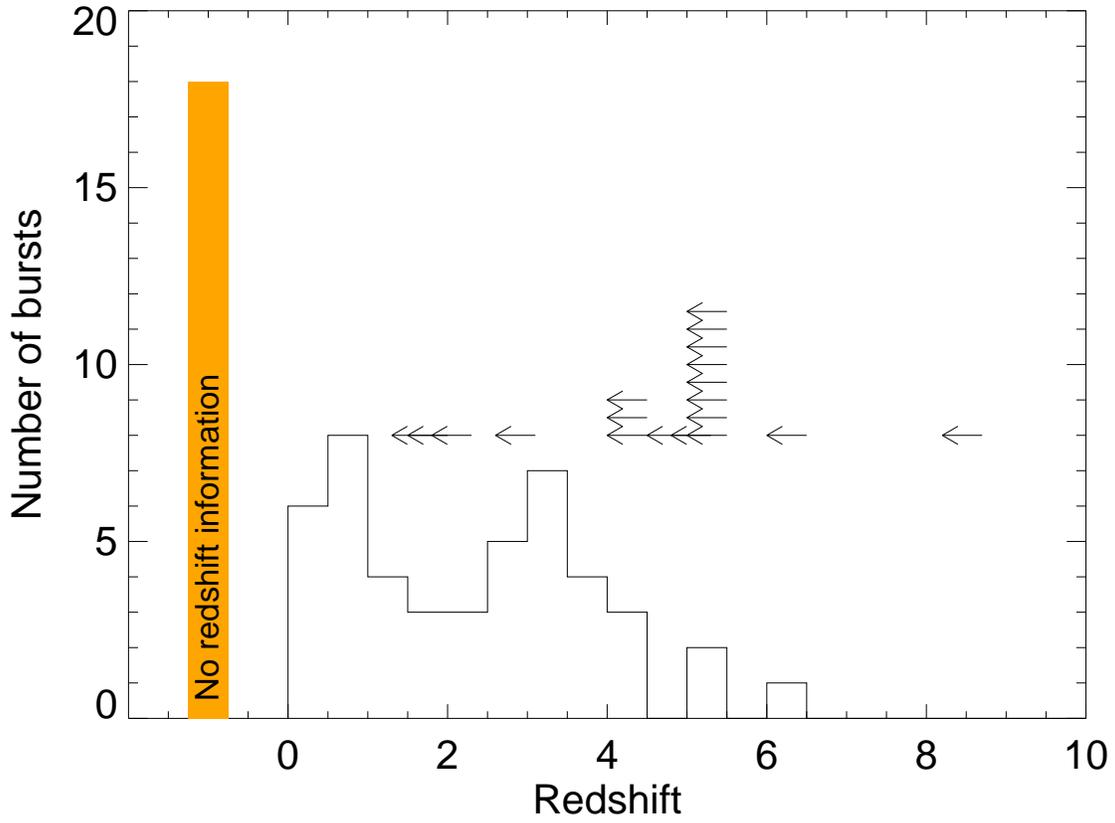}
\caption{The redshift distribution of \textit{Swift} GRBs up to 2006 December.
The arrows indicate upper limits based on the afterglow colors and/or presence of
Ly$\alpha$ breaks in the spectra and/or detection of the host galaxy. The
shaded bar indicates the 18 bursts for which we have no constraints on the redshift.
\label{fg:zdist}}
\end{figure}

\textit{Swift} has been operating for more than two years and we now have a
good sample of GRBs from which to constrain their fraction at very high
redshift. We followed \citet{palli06} and selected only \textit{Swift} GRBs
with low Galactic extinction and promptly determined XRT positions. In this
way we created an unbiased sample with a larger redshift completeness, discarding those bursts with adverse
observing conditions. By December 2006, there had been a total of 83 of these
bursts: 46 with spectroscopic redshift measurements (with a mean
$\langle z \rangle = 2.31$), 18 with no optical/NIR afterglow nor host galaxy detection, and 19
with optical/NIR afterglow or host galaxy detection but unknown spectroscopic redshift.
In order to compile this list, we browsed the literature, and we also
made use of the preliminary results of an ongoing large program aimed at
detecting GRB host galaxies and measuring their redshifts
(J. Hjorth et al. 2007 in preparation; \citet{palli06c}).
In Fig.~\ref{fg:zdist} we show the redshift distribution for the 83 bursts. 
The arrows indicate upper limits based on the information available:
afterglow colors, host galaxy and/or Ly$\alpha$ break constraints.
The shaded bar indicates bursts for which we have no redshift information at all.

Several bursts in the sample with very red afterglows have shown properties
expected for high-redshift bursts: GRB\,050502B, GRB\,050716, GRB\,060108,
GRB\,060602A, GRB\,060719, GRB\,060814, GRB\,060923A and GRB\,060923C 
\citep[][respectively]{Cenko05,Tanvir05,Levan06a,Jensen,Malesani06,Levan06b,Fox,DAvanzo}.
However, for GRB\,060108 \citep{Oates06b}, GRB\,060814 \citep{Campana06}, and
GRB\,060923A \citep{Levan06c}, the detection of the host galaxy in the $R$ band
indicates $z \lesssim 5.5$ and hence the red afterglow color is due
to dust extinction rather than to high redshift%
\footnote{The afterglow of GRB\,060108 was also detected in the $B$ band,
implying a low redshift $z < 3.2$ \citep{Oates06b}.}.

We now try to constrain the fraction $f$ of \textit{Swift} GRBs at $z > 6$.
From the initial sample we removed 
bursts with measured or constrained redshift, leading to a sub-sample of 20 candidates: 
18 bursts with no information at all and
2 bursts with very loose redshift constraints. We rejected from this set those
bursts with duration shorter than 14~seconds\footnote{A burst with $T_{90} = 14$~s at $z=6$ has
a proper time duration of $\approx2$~s; see also \citet{Campana06}.}
and those with (significant) excess column density in the X-ray spectra. As
pointed out by \citet{grupe}, high-redshift GRBs tend to show little or no X-ray
absorption, since the affected spectral region is redshifted out of the
observing window (see however the striking counterexample of GRB\,050904 at
$z=6.295$: \citealt{Watson06,Cusumano06,Campana07}). Only six bursts 
were considered as candidates for being at $z>6$. Together with GRB\,050904, they set an upper
limit of $f <7/83 \approx 8\%$. 
Do we have any evidence that these bursts are at
high redshift?
We retrieved from the literature optical magnitude upper
limits\footnote{\texttt{http://grad40.as.utexas.edu/grblog.php}\,.} and X-ray
fluxes\footnote{\texttt{http://astro.berkeley.edu/$\sim$nat/swift}\,.}
\citep{ButlerKocevski07} in order
to estimate an upper limit to the optical-to-X-ray spectral index $\beta_{\rm
OX}$. High-redshift bursts necessarily have $\beta_{\rm OX} < 0.5$ \citep{palli04}.
Unfortunately, the existing observational limits are not deep enough to
constrain $\beta_{\rm OX} < 0.5$ for any of the candidates. By considering
GRB\,050904, we can thus set a lower limit $f \gtrsim 1/83 \approx 1\%$, 
even if we
caution, of course, that this limit is based on one single event. It has been
argued \citep{ButlerKocevski07} that afterglow spectra may be intrinsically
curved (due to the presence of spectral breaks) and this could spuriously
mimic the presence of excess column density. If we remove the requirement of zero
excess absorption, the upper limit on $f$ becomes $f < 16/83 \approx 19\%$, 
which is of course less constraining and very conservative.

From what is shown above, it is possible that \textit{Swift} had
detected some GRBs with $z > 6$ in addition to GRB~050904.
How do we improve our follow-up
strategies in order to identify these bursts and secure spectroscopic
redshifts? It is clear from the discussion above that the afterglows of such
events will be very red due to Lyman-forest blanketing, but that not all red
afterglows are from high redshift bursts. Fortunately, the dust-extinguished bursts
will likely be behind large column densities of soft X-ray absorbing material
and hence can be filtered away \citep{grupe}. The main issue is therefore to
detect the NIR afterglows.

The two most distant GRBs known, GRB\,050904
\citep[e.g.][]{Kawai,haislip,gendre,totani06,Taglia05} and GRB\,060927 (this work),
were both detected with relatively small aperture robotic telescopes (TAROT and
ROTSE-IIIa, respectively). This suggests that we may expect bright afterglows
from a substantial fraction of GRBs with $z>6$, but these will not be
detectable with most of the currently operating robotic instruments because of
the lack of light in the optical window. 
NIR robotic projects (e.g. REM, GROND on the ESO 2.2m, the 1.3m CTIO
telescope, PAIRITEL, BOOTES-IR) are expected to provide accurate afterglow
positions in just a few minutes, and, hence, to increase the sample of
spectroscopically measured high-redshift GRBs. However, three of these are
placed within a few hundred kilometers of each other in Chile. In order to make
the search of events that are only visible in the NIR as efficient as in the
optical, a more wide-spread net of NIR (semi)-robotic telescopes is required. A
possible solution, easy to implement on a short time scale, is to equip many of
the 1.5--3~m class telescopes, that have had less observing demands since the
advent of 8--10~m telescopes, with red-sensitive deep-depletion CCDs and
$z$-band filters \citep[see e.g.][]{guidorzib}. In this way the probability of
detecting bursts out to $z \approx 7.5$ would dramatically increase.

\section{SUMMARY AND CONCLUSIONS}

We have presented X-ray and optical observations of the afterglow of
GRB~060927, performed with the \textit{Swift} spacecraft and several
ground-based telescopes. The spectroscopic observations made with the VLT
showed a flux dropout starting at $\lambda = 8070$~\AA. We interpret the
absorption blueward of the break as due to the Ly$\alpha$ forest at
redshift $z \approx 5.6$. We also detect a \ion{Si}{2} $\lambda$ 1260
absorption line at $\lambda = 8148$~\AA{}, corresponding to $z = 5.467$.

The modeling of the red damping wing of the Ly$\alpha$ absorption found in
the optical afterglow spectrum leads to a column density with $\log(N_{\rm
HI} / {\rm cm}^{-2}) = 22.50 \pm 0.15$. This is similar to the values observed in
GRBs at lower redshift. Such large values make the use of the GRBs to
probe the reionization epoch via spectroscopy of the red damping wing
challenging.

The light curve in the gamma-ray band shows an initial bright episode,
split in two peaks, with a total duration $T_{90} = 22.6 \pm 0.3$~s. The X-ray
light curve is characterized by a shallow decline phase lasting $\sim
3000$~s, with the temporal index being $\alpha_1 \approx 0.7$. 
The shallow decline steepened to $\alpha_2
\approx 1.3$ at $\sim 4$~ks after the trigger. The optical data are
fairly well described by a single power law with decay index $\alpha =
1.17\pm 0.03$ after $\sim 500$~s, while an extra component is apparent at
earlier times.

We selected a sample of \textit{Swift} GRBs with low Galactic
extinction and promptly determined XRT positions. From this unbiased
sample, we constrain the fraction $f$ of GRBs at $z > 6$. Including
GRB\,050904, we set a conservative upper limit $f<19\%$. We also point
out that GRB\,060927 was detected by a 45 cm telescope (ROTSE). This shows
that high-redshift GRBs can be very bright, and are in principle easy to
detect, provided that red and near-infrared detectors are employed. We
suggest to exploit the potential of the $z$-band filter, capable to detect
bursts up to $z \approx 7.5$.

\acknowledgments

The Dark Cosmology Centre is funded by the Danish National Research Foundation.
The authors acknowledge benefits from collaboration within the EU FP5 Research
Training Network ``Gamma-Ray Bursts: An Enigma and a Tool''. DM acknowledges
financial support from the Instrument Center for Danish Astrophysics. JG
acknowledges the support of Spanish research programmes ESP2005-07714-C03-03
and AYA2004-01515. PJ acknowledges support by a Marie Curie Intra-European Fellowship
within the 6th European Community Framework Program under contract
number MEIF-CT-2006-042001. ROTSE-III has been supported by NASA grant NNG-04WC41G, NSF
grant AST-0407061, the Australian Research Council, the University of New South
Wales, the University of Texas, and the University of Michigan. Work performed
at LANL is supported through internal LDRD funding. Special thanks to the
observatory staff at Paranal and Siding Spring Observatory, especially Andre
Phillips.

\clearpage


\begin{deluxetable}{lllllll}
\tabletypesize{\scriptsize}
\tablewidth{0pt}
\tablecaption{Ground-based observations of GRB\,060927. Upper limits are at the 3$\sigma$ level.\label{tb:log}}
\tablehead{
\colhead{Time (UT)\tablenotemark{a}} & 
\colhead{$\Delta t$ (s)} & 
\colhead{Exposure (s)} & 
\colhead{Magnitude\tablenotemark{b}} & 
\colhead{$F_\nu(\lambda_0)$\tablenotemark{c}} &
\colhead{Filter} & 
\colhead{Telescope}
}
\startdata
Sep. 27.588794 & 16.8               & 5             & 14.39$\pm$0.20  & $6800\pm1300$  & none        & 0.45-m ROTSE-IIIa   \\
Sep. 27.588956 & 30.8               & 5             & 14.67$\pm$0.22  & $5200\pm1100$  & none        & 0.45-m ROTSE-IIIa   \\
Sep. 27.589216 & 53.3               & 15            & 15.56$\pm$0.19  & $2290\pm410$   & none        & 0.45-m ROTSE-IIIa   \\
Sep. 27.589804 & 104.1              & 25            & 17.16$\pm$0.41  & $520\pm200$    & none        & 0.45-m ROTSE-IIIa   \\
Sep. 27.590624 & 174.9              & 200           & 17.86$\pm$0.30  & $275\pm78$     & none        & 0.45-m ROTSE-IIIa   \\
Sep. 27.594007 & 467.2              & 600           & 17.07$\pm$0.21  & $570\pm110$    & none        & 0.45-m ROTSE-IIIa   \\
Sep. 27.602126 & 1168.7             & 600           & 17.99$\pm$0.40  & $243\pm92$     & none        & 0.45-m ROTSE-IIIa   \\
Sep. 27.654894 & 5727.9             & 600           & $>17.04$        & $< 590$        & none        & 0.45-m ROTSE-IIIa   \\
Sep. 27.593845 & 453.2              & 930           & $>20.1$         & \nodata        & SDSS $R$    & 2.0-m FTS           \\
Sep. 27.597834 & 797.9              & 60            & $>18.4$         & $< 570$        & SDSS $i$    & 2.0-m FTS           \\
Sep. 27.602725 & 1221.3             & 120           & 19.02$\pm$0.17  & $320\pm50$     & SDSS $i$    & 2.0-m FTS           \\
Sep. 27.609716 & 1824.4             & 180           & 19.64$\pm$0.22  & $180\pm37$     & SDSS $i$    & 2.0-m FTS           \\
Sep. 27.616009 & 2368.2             & 120           & 19.43$\pm$0.27  & $220\pm55$     & SDSS $i$    & 2.0-m FTS           \\
Sep. 27.623009 & 2973.0             & 180           & 20.57$\pm$0.47  & $77\pm33$      & SDSS $i$    & 2.0-m FTS           \\
Sep. 27.602650 & 1214.0             & 90            & 19.5$\pm$0.3    & \nodata        & $R$         & 1.05-m Kiso         \\
Sep. 27.627431 & 3355.0             & 90            & 19.4$\pm$0.1    & \nodata        & $R$         & 1.05-m Kiso         \\
Sep. 27.629468 & 3531.0             & 90            & 19.8$\pm$0.2    & \nodata        & $R$         & 1.05-m Kiso         \\
Sep. 27.844120 & $2.21\times10^{4}$ & 3$\times$300  & 24.4$\pm$0.3    & \nodata        & $R$         & 2.5-m NOT           \\
Sep. 27.957326 & $3.18\times10^{4}$ & 4$\times$100  & $>25.5$         & \nodata        & $B$         & 4.2-m WHT           \\
Sep. 27.964109 & $3.24\times10^{4}$ & 4$\times$100  & $>24.6$         & \nodata        & $V$         & 4.2-m WHT           \\
Sep. 27.973495 & $3.32\times10^{4}$ & 4$\times$100  & $>24.2$         & \nodata        & $R$         & 4.2-m WHT           \\
Sep. 27.978808 & $3.37\times10^{4}$ & 4$\times$100  & 22.11$\pm$0.16  & $7.2\pm1.1$    & Harris $I$  & 4.2-m WHT           \\
Sep. 28.006500 & $3.61\times10^{4}$ & 13$\times$900 & 24.5$\pm$0.26   & \nodata        & Bessell $R$ & 1.54-m Danish       \\
Sep. 28.064850 & $4.11\times10^{4}$ & 2$\times$60   & $>$24.0         & \nodata        & Bessell $R$ & 8.2-m VLT           \\
Sep. 28.108681 & $4.49\times10^{4}$ & 3$\times$1800 & \nodata         & \nodata        & G300V       & 8.2-m VLT           \\
Sep. 30.004618 & $2.09\times10^{5}$ & 10$\times$30  & 21.35$\pm$0.29  & \nodata        & $K$         & 8.2-m VLT           \\
Sep. 30.039178 & $2.12\times10^{5}$ & 30$\times$60  & 23.15$\pm$0.45  & \nodata        & $J$         & 8.2-m VLT           \\
Sep. 30.159225 & $2.22\times10^{5}$ & 21$\times$300 & 24.85$\pm$0.17  & $0.72\pm0.11$  & Bessell $I$ & 8.2-m VLT           \\
Oct. 13.067980 & $1.34\times10^{6}$ & 25$\times$300 & $>25.8$         & $< 0.3$        & Bessell $I$ & 8.2-m VLT           \\
\enddata
\tablenotetext{a}{The reported times correspond to the beginning of the observation.}
\tablenotetext{b}{Not corrected for Galactic extinction.}
\tablenotetext{c}{Flux density at $\lambda_0 = 8190$~\AA, measured in $\mu$Jy, corrected for Galactic extinction $E(B-V) = 0.06$~mag.}
\end{deluxetable}

\begin{deluxetable}{llllllll}
\tabletypesize{\scriptsize}
\tablecaption{Model fits to the SED at epoch 5709 s since trigger. All errors are at the 90\% confidence level.
\label{tb:SED}}
\tablewidth{0pt}
\tablehead{
\colhead{Model\tablenotemark{a}} &
\colhead{Ext. curve\tablenotemark{b}} &
\colhead{$N_{\rm H}$ \tablenotemark{c}} &
\colhead{$E(B-V)$} &
\colhead{$\Gamma_1$}&
\colhead{$E_{\rm b}$ (keV)} &
\colhead{$\Gamma_2$} &
\colhead{$\chi^2$/dof}
}
\startdata
PL                            &MW  &$< 1.2$     &0.13$^{+0.03}_{-0.02}$ &1.70$\pm$0.06          &\nodata &\nodata                 &9.5/20  \\
PL                            &MW  &0 (frozen)  &0.13$^{+0.06}_{-0.05}$ &1.70$\pm$0.06          &\nodata &\nodata                 &9.5/21  \\
PL                            &LMC &$< 1.1$     &0.08$^{+0.02}_{-0.01}$ &1.68$^{+0.04}_{-0.06}$ &\nodata &\nodata                 &10.7/20 \\
PL                            &LMC &0 (frozen)  &0.08$^{+0.04}_{-0.03}$ &1.68$^{+0.05}_{-0.06}$ &\nodata &\nodata                 &10.7/21 \\
PL                            &SMC &$< 1.0$     &0.07$^{+0.01}_{-0.02}$ &1.61$^{+0.05}_{-0.04}$ &\nodata &\nodata                 &11.5/20 \\
PL                            &SMC &0 (frozen)  &0.07$\pm$0.03          &1.66$^{+0.10}_{-0.05}$ &\nodata &\nodata                 &11.5/21 \\
BPL                           &MW  &$< 3.6$     &0.11$^{+0.06}_{-0.07}$ &1.64$^{+0.09}_{-0.21}$ &$<$106  &1.83$^{+7.17}_{-0}$     &6.6/18  \\
BPL ($\Gamma_1=\Gamma_2-0.5$) &MW  &$< 4.7$     &0.13$^{+0.06}_{-0.09}$ &                       &$>$1.0  & 2.19$^{+0.06}_{-0.13}$ &8.9/19  \\
BPL                           &LMC &$< 3.6$     &0.07$\pm$0.04          &1.63$^{+0.07}_{-0.21}$ &$<$2.7  &1.83$^{+0.33}_{-0}$     &7.0/18  \\
BPL ($\Gamma_1=\Gamma_2-0.5$) &LMC &$< 4.8$     &0.07$^{+0.05}_{-0.04}$ &                       &$>$0.9  &2.13$^{+0.09}_{-0.08}$  &9.5/19  \\
BPL                           &SMC &$< 3.5$     &0.05$^{+0.04}_{-0.03}$ &1.66$^{+0.07}_{-0.21}$ &$<$2.4  &1.83$^{+0.31}_{-0}$     &7.0/18  \\
BPL ($\Gamma_1=\Gamma_2-0.5$) &SMC &$< 4.8$     &0.05$^{+0.04}_{-0.03}$ &                       &$>$0.8  &2.12$^{+0.07}_{-0.08}$  &9.5/19  \\
\enddata
\tablenotetext{a}{PL: power law with photon index $\Gamma_1$; BPL: broken power law with photon indices $\Gamma_1$, $\Gamma_2$ and break energy $E_{\rm b}$.}
\tablenotetext{b}{MW: Milky Way; LMC: Large Magellanic Cloud; SMC: Small Magellanic Cloud.}
\tablenotetext{c}{Rest-frame hydrogen column density in units of $10^{22}$~cm$^{-2}$.}
\end{deluxetable}

\clearpage


\clearpage

\clearpage

\clearpage

\clearpage

\clearpage

\end{document}